\documentclass[mathleft
]{an}
\pdfoutput=1
\usepackage{graphicx}
\usepackage{times}
\usepackage{natbib}
\bibpunct{(}{)}{;}{a}{}{,}
\def\note #1]{{\bf #1]}}
\def\trl{png}     
\begin{document}

\Pagespan{1088}{1091}
\Yearpublication{2012}%
\Yearsubmission{2012}%
\Month{10}%
\Volume{333}%
\Issue{10}%
\DOI{10.1002/asna.201211832}

\title{Properties of extrasolar planets and their host stars -- \\
        a case study of HAT-P-7}

\author{V. Van Eylen\inst{1,2}\fnmsep\thanks{Corresponding author:
  \email{vincent.vaneylen@gmail.com}\newline}
\and  H. Kjeldsen\inst{1}
\and J. Christensen-Dalsgaard\inst{1}
\and C. Aerts\inst{2}
}
\titlerunning{Properties of extrasolar planets and their host stars}
\authorrunning{V. Van Eylen et al.}
\institute{
Stellar Astrophysics Centre, Department of Physics and Astronomy, Aarhus University, Ny Munkegade 120, \\
DK-8000 Aarhus C, Denmark
\and
Instituut voor Sterrenkunde, Katholieke Universiteit Leuven, Celestijnenlaan 200 B, B-3001 Heverlee, Belgium
}

\received{15 August 2012}
\accepted{15 August 2012}

\publonline{3 December 2012}

\keywords{stars: individual (HAT-P-7) -- stars: oscillations -- planetary systems -- stars: fundamental parameters}

\abstract{%
Data from the \textit{Kepler} satellite (Q0-Q11) are used to study HAT-P-7. The satellite's data are extremely valuable for asteroseismic studies of stars and for observing planetary transits; in this work we do both. An asteroseismic study of the host star improves the accuracy of the stellar parameters derived by \cite{hatp7joergen}, who followed largely the same procedure but based the analysis on only one month of \textit{Kepler} data. The stellar information is combined with transit observations, phase variations and occultations to derive planetary parameters. In particular, we confirm the presence of ellipsoidal variations as discovered by \cite{hatp7ellipsoidalvariations}, but revise their magnitude, and we revise the occultation depth \citep{hatp7science}, which leads to different planetary temperature estimates. All other stellar and planetary parameters are now more accurately determined.
}	

\maketitle

\section{Introduction}

The \textit{Kepler} mission was launched on March 6, 2009. It was designed explicitly to be able to detect earth-size planets, including those in the habitable zone. The detection method for \textit{Kepler} to achieve this goal is measuring planetary transits to a high level of photometric precision. To improve the chances of detection, a large number of stars have to be monitored for a long period of time, preferably continuously. It was quickly realised that \textit{Kepler} would also provide extremely valuable data to study stellar pulsations. Data for this purpose are obtained from a subset of the about 500 stars that are sampled every 58.8 seconds (short cadence), out of a much larger set of about 150 000 stars observed at a rate of 29.4 minutes \citep{borucki2008}.

{
The satellite orbits in an earth-trailing heliocentric orbit (ETHO), with a 55$^\circ$ Sun avoidance angle, and continuously observes a patch of the sky centred on the Cygnus-Lyra region. The data are organised into quarters; after each quarter the photometer is rolled 90$^\circ$ to keep the solar arrays pointed at the Sun. This causes a 42 hour data gap \citep{haas2010}. This work uses Q0-Q11 data of HAT-P-7, which has been continuously observed in \textit{Kepler}'s short cadence mode.
}

HAT-P-7 (KIC 10666592, visual magnitude 10.46, spectral type F8) has been identified as a planet host by ground-based observations prior to the start of the \textit{Kepler} mission \citep{pal2008}. It is one of the most extensively studied star-planet systems known to date. A study of the stellar pulsations has been carried out by \cite{hatp7joergen}, based on early \textit{Kepler} data (Q0-Q1). After the initial discovery of its exoplanet HAT-P-7b \citep{pal2008}, phase variations and planetary occultations were discovered in \textit{Kepler}'s Q0 data \citep{hatp7science}, and consequently the planetary temperature and atmosphere were studied \citep{hatp7hotdayside,hatp7atmosphere}. Additional ellipsoidal variations were later discovered in the light curve \citep{hatp7ellipsoidalvariations} and the system is currently the only planetary case where this effect has been found. HAT-P-7 has a surprising inclination angle: a study of the Rossiter-McLaughlin effect in radial velocity data has found it to be 
retrograde or polar \citep{hatp7thirdbody}. These authors also suggest the presence of a third body. A study of transit timings based on EPOXI data found no evidence of an additional planet \citep{hatp7additionalplanets}. 

Here, we make a preliminary anaysis of the system based on almost three years of \textit{Kepler} data. We derive stellar pulsation frequencies and perform an asteroseismic modelling to derive the properties of the host star. The planetary system is then studied making use of these newly determined stellar properties. 

\section{Stellar properties}

Stellar pulsations cause temperature variations, which lead to flux variations. The frequencies are observed from the power spectrum. Different models are then compared with the observed frequencies and the best fit leads to the stellar properties. The procedure followed here resembles the asteroseismic analysis of HAT-P-7 based on only the first month of \textit{Kepler} data \citep{hatp7joergen}.

\subsection{Observations and data analysis}

The raw \textit{Kepler} data are detrended and normalised, and transit-like features are removed using a specifically designed median filter. The filtered photometry is then used to calculate a power spectrum. A spectrum with a sampling rate of 0.005 $\mu$Hz is used, which means an oversampling of about three times. A median filter with a period of 1 $\mu$Hz is used to smooth the power spectrum, to remove the fine structure of the spectrum, caused by the finite lifetime of the modes.

\begin{figure}
\includegraphics[width=1\columnwidth]{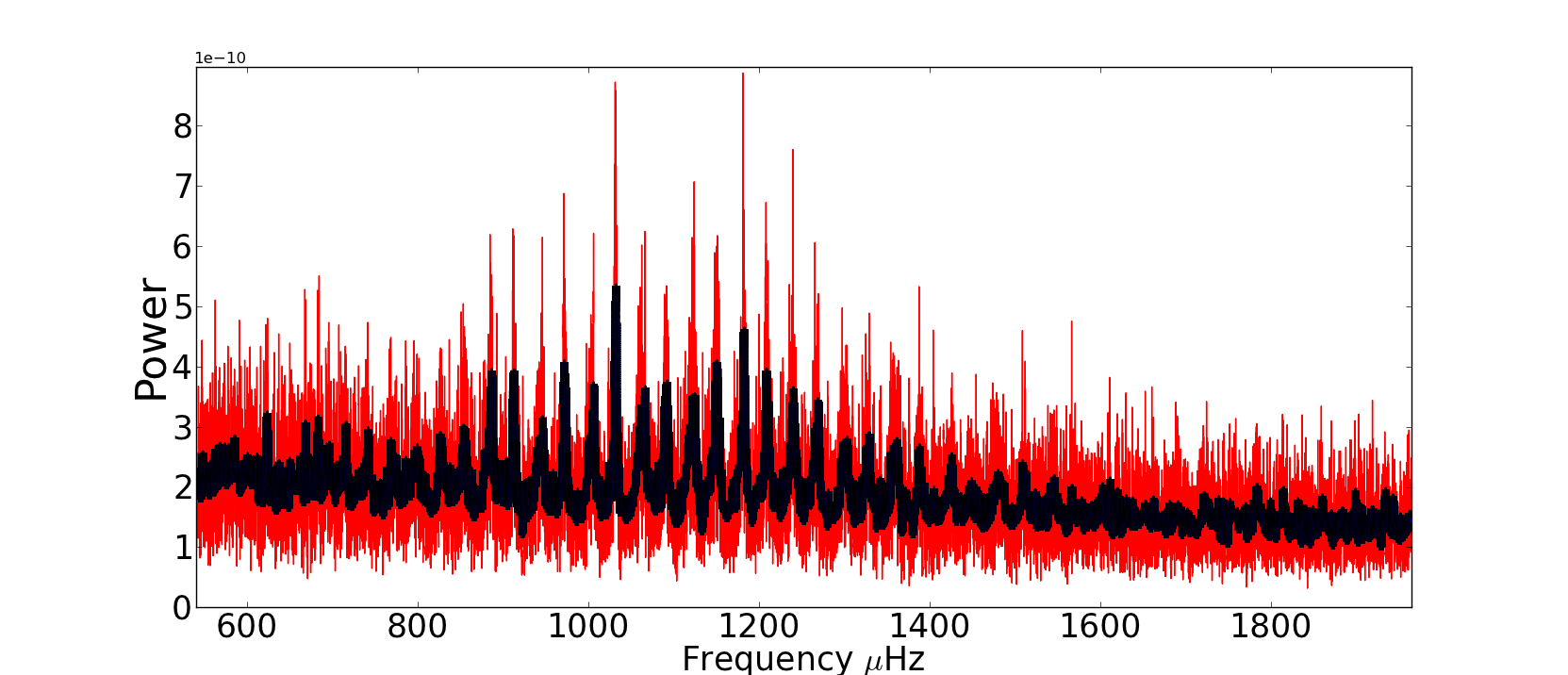}
\includegraphics[width=1\columnwidth]{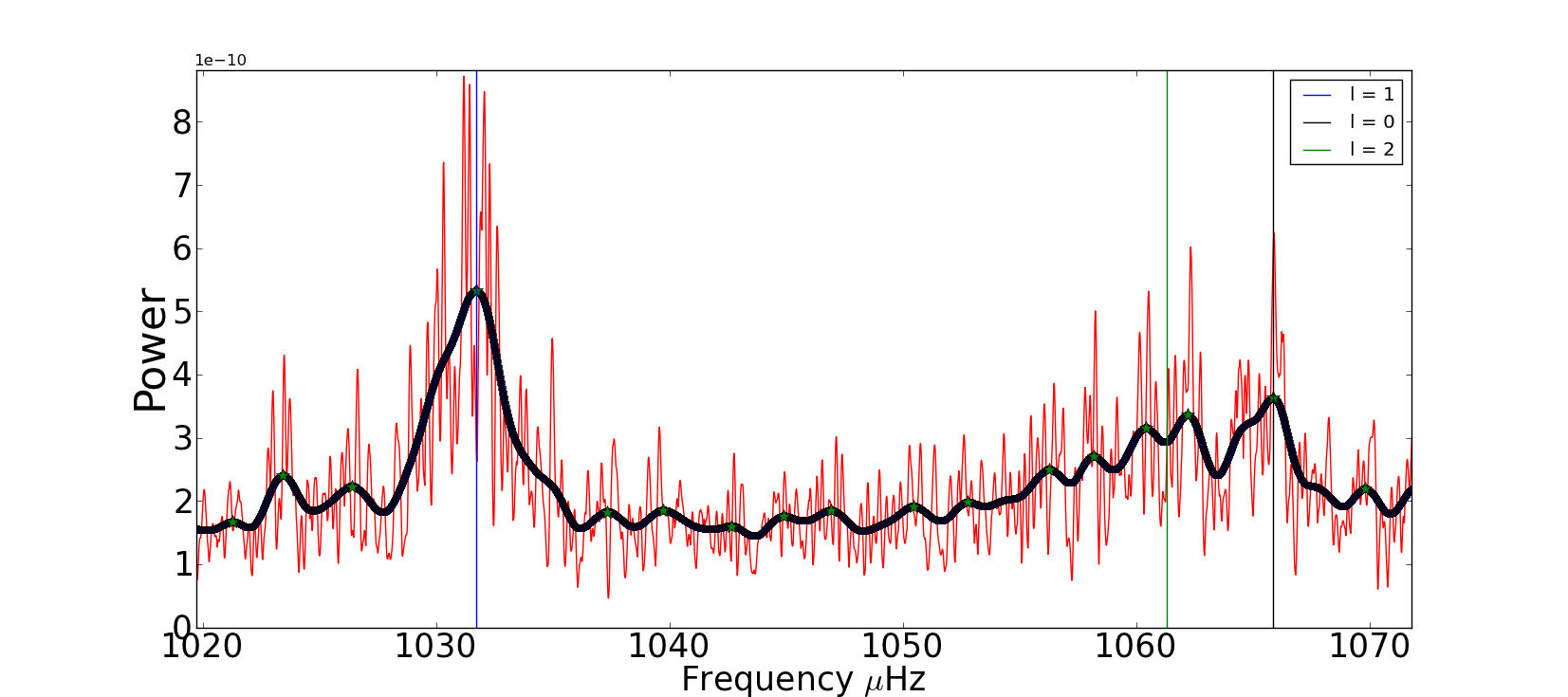}
\caption{Top figure: oversampled (red) and smoothed power spectrum (black), showing regular spacings governed by the asymptotic equation (eq. \ref{eq:asymptotic}). Bottom figure (zoom): mode identification for three observed frequencies. An algorithm determines the maxima in the smoothed spectrum, the relevant maxima (vertical lines) are selected manually making use of the asymptotic equation. Sometimes, an average of two maxima is used (e.g. $l = 2$ mode in this figure).}
\label{frequency_observations}
\end{figure}

The location of individual frequencies is governed by the asymptotic equation \citep{tassoul1980}:

\begin{equation}\label{eq:asymptotic}
 \nu_{n,l} \approx \Delta \nu \left( n + \frac{1}{2}l + \epsilon \right) - \delta \nu_{0l}.
\end{equation}
In this equation, $\nu_{n,l}$ is the frequency with radial order $n$ and angular degree $l$ (neglecting rotational splitting so there is no dependence on $m$), and $\Delta \nu$ and $\delta \nu_{0l}$ the large and small separations while $\epsilon$ is an offset coming from second order and near-surface effects. This theoretical relation is used to identify the individual modes, from the smoothed power spectrum. The entire procedure is illustrated in Figure \ref{frequency_observations}.

We have determined 14 ($l = 0$), 13 ($l = 1$) and 16 ($l = 2$) frequencies, which constitute a total of ten more frequencies than the initial asteroseismic analysis \citep{hatp7joergen}. The frequencies were determined with an average standard error estimated to be $\sigma_\nu = $ 0.75 $\mu$Hz (compared to a previous accuracy of 1.4 $\mu$Hz). They are illustrated in Figure~\ref{fig:echelle}.

\subsection{Modelling}

To calculate evolution models and adiabatic oscillation frequencies, we use the Aarhus codes \citep{ASTEC,ADIPLS}. Evolution tracks were calculated for a grid of models with mass range 1.33-1.70 M$_\odot$ and metallicity 0.015-0.035. The convective core was modelled with overshoot in units of the pressure scale height of $\alpha_{\textrm{ov}} = 0$, 0.1 and 0.2. Based on the observed large separation an estimate of the density is obtained \citep{temperaturescalingrelation}, which is used to select relevant time steps in the stellar evolution tracks, for which frequencies are then calculated.

The calculated frequencies are compared with the observations, after correcting for near-surface effects that are not modelled properly. The empirical correction law is given by \citep{surfacecorrection}:

\begin{equation}
 \nu_{\rm{obs}}(n) - \nu_{\rm{best}}(n) = a \left[\frac{\nu_{\rm{obs}}(n)}{\nu_0}\right]^b,
\end{equation}
where $\nu_0$ is a reference frequency and $a$ and $b$ are parameters to be determined. In addition, we set $\nu_{\rm{best}}(n) = r\nu_{\rm{ref}}(n)$, where $r$ should be close to one as the reference model should be close to the initial best model. In practice, we set $b = 4.90$ (the solar value) and we find the best fit for $r$ and $a$ (for each model individually). We can then calculate the reference model, for comparison with the observed frequencies. The best model is then picked by comparing the observed frequencies with the corrected model frequencies. We use a $\chi^2$ minimisation to determine which model fits best:

\begin{equation}
 \chi^2 = \frac{1}{(N-1)\sigma_\nu^2} \sum_{nl} \left( \nu_{\rm{nl}}^{\rm{(obs)}} - \nu_{\rm{nl}}^{\rm{(mod)}} \right)^2.
\end{equation}

\begin{figure}
\includegraphics[width=1\columnwidth]{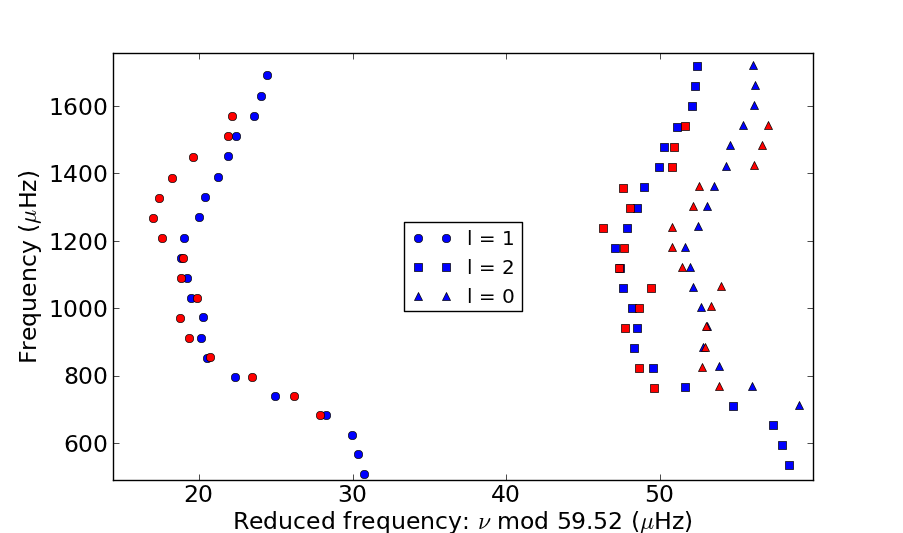}
\caption{An \'echelle diagram showing the observed frequencies (red) and the best model (blue).}
\label{fig:echelle}
\end{figure}

The frequencies of the best model are shown together with the observational frequencies in the \'echelle diagram in Figure~\ref{fig:echelle}. With the exception of a number of frequencies (between 1200 and 1400 $\mu$Hz) for $l = 1$, the frequencies are modelled very well, resulting in $\chi^2 = 3.65$. Gaussian noise was added to the observational frequencies to create 1000 different frequency sets. For each, the best model is calculated and the stellar parameters are calculated. The average and standard deviation of these parameters are presented in Table~\ref{tab:hatp7_parameterlist}.

\begin{table}
 \caption{Parameters obtained for star (HAT-P-7) and its planet (HAT-P-7b).
\label{tab:hatp7_parameterlist}}
\begin{center}
\begin{tabular}[width=1\columnwidth]{ccc}
\hline
  \multicolumn{2}{c}{HAT-P-7}  \\
\hline
  Mass (M$_\odot$) 		& 1.361 $\pm$ 0.021 		\\
  Z$_0$				& 0.01904 $\pm$ 0.0015 		\\
  $[\rm{Fe}/\rm{H}]$		& 0.13				\\
  Age (Gyr) 			& 2.19 $\pm$ 0.12 		\\
  Radius (R$_\odot$) 		& 1.904 $\pm$ 0.010 		\\
  Temperature (K) 		& 6259 $\pm$ 32 		\\
  Density (g/cm$^{-3}$) 	& 0.2781 $\pm$ 0.0017		\\
  Luminosity (L$_\odot$) 	& 4.996 $\pm$ 0.098		\\
  $\log g$ (cgs) 		& 4.012 $\pm$ 0.008		\\
  Ellips. var. $d$ (ppm)	& 59 $\pm$ 1			\\
\hline
\hline
  \multicolumn{2}{c}{HAT-P-7b}  \\
\hline
Orbital period (d) 			& 2.20473506	(11)	\\
Orbital inclination i ($^\circ$) 	& 86.68 $\pm$ 0.14	\\
$M_{\rm p}$ (M$_J$) 			& 1.741 $\pm$ 0.028	\\

$R_{\rm p}$ (R$_J$)			& 1.431 $\pm$ 0.011	\\
$T_{\rm{day}}$	(K)			& 2470 $\pm$ 10 	\\
$T_{\rm{night}}$ (K)			& 1772 $\pm$ 10		\\
$A_{\rm{g}}$ 				& 0.193 $\pm$ 0.002	\\
\hline
\end{tabular}
\end{center}
\end{table}

\section{Planetary properties}

\subsection{Period}

The time of each individual transit is measured and used to determine the planetary period. Assuming a perfectly Keplerian orbit, we fit a straight line through all transit times (taking gaps into account; whenever a transit is measured incompletely it is ignored). Due to the large amount of measured transits (286), the period is determined very accurately. An O-C plot has been calculated, which compares the observed and the calculated transit times to look for a gravitational influence from additional bodies in the system. No sinuisoidal pattern was detected, and variations are at the level of the observational error level for individual transit times ($\approx 0.00015$ days). After the period has been calculated, the time series is folded (calculated modulo the period).

\subsection{Transit shape}

From the folded time series, the transit is modelled. The transit duration determines the impact parameter, which is directly related to the orbital inclination. The transit depth determines the relative planetary radius, which can be combined with the stellar radius (derived in the previous section), to obtain the absolute planetary radius. The shape of the bottom of the transit depends on the limb darkening, modelled by a  quadratic intensity profile \citep{exoplanetsbook_chapterwinn}:

\begin{equation}
I(X,Y) \propto 1 - u_1 (1-\mu) - u_2 (1-\mu)^2,
\end{equation}
with $\mu = \sqrt{1 - X^2 - Y^2}$ and $(X,Y)$ the coordinates on a unit circle. The coefficients $u_k$ are constants defining the precise shape of the limb-darkening law. We have simply used them as fitting parameters. The modelled transit is shown in Figure~\ref{fig:folded_flux}. We have also made a new estimate of the planetary mass, using the radial velocity values derived by \cite{pal2008}.

\subsection{Phase variations}

The planetary system is tidally locked, which causes the star to always `see' the same side of the planet, while a distant observer sees the planetary dayside and nightside at different times. However, for HAT-P-7 the situation is slightly complicated. The close orbit of the planet has induced a tidal distortion on the star: rather than being oblate due to rotation, it has its longest axis towards the planet and the shortest axis perpendicular to the orbital plane. This causes ellipsoidal variations (which peak at phases near 0.25 and 0.75), which should be added to the phase variations \citep{pfahl2008,hatp7ellipsoidalvariations}. HAT-P-7 is the only currently known system with ellipsoidal variations caused by a planet. We model the flux throughout one orbital phase as

\begin{eqnarray}
f(\Phi, i) &=& A_{\rm g} \left(\frac{R_{\rm p}}{a}\right)^2 \frac{\sin(\alpha) + (\pi-\alpha)\cos(\alpha)}{\pi}\\
 & & - d \frac{\cos(2\alpha)}{\pi} \nonumber
\end{eqnarray}
%
with the orbital phase $\Phi \in$ [0,1] taken as 0 at maximum radial velocity of the star, so that the phase angle $\alpha$ is defined as $\cos \alpha = - \sin i \sin 2\pi \Phi.$ The first part of $f(\Phi, i)$ is the Lambert law for a sphere \citep{planetaryreflection}, where $A_{\rm g}$ is the planetary albedo and $a$ is the semi-major axis. A non-zero value for $d$ corresponds to the ellipsoidal variations. We refer to Table~\ref{tab:hatp7_parameterlist} for values and to Figure~\ref{fig:folded_flux} for the best fit.

\subsection{Occultation}

\begin{figure}
\includegraphics[width=1\columnwidth]{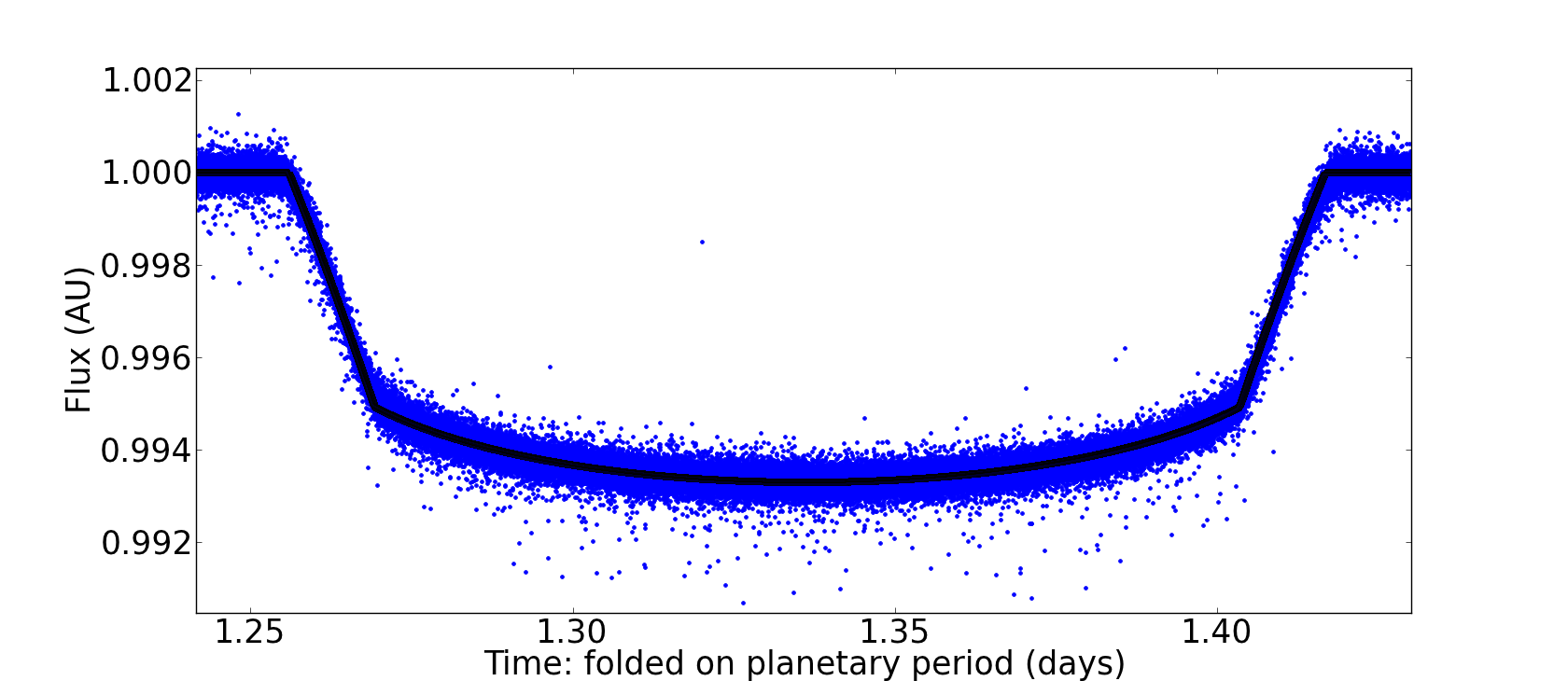}
\includegraphics[width=1\columnwidth]{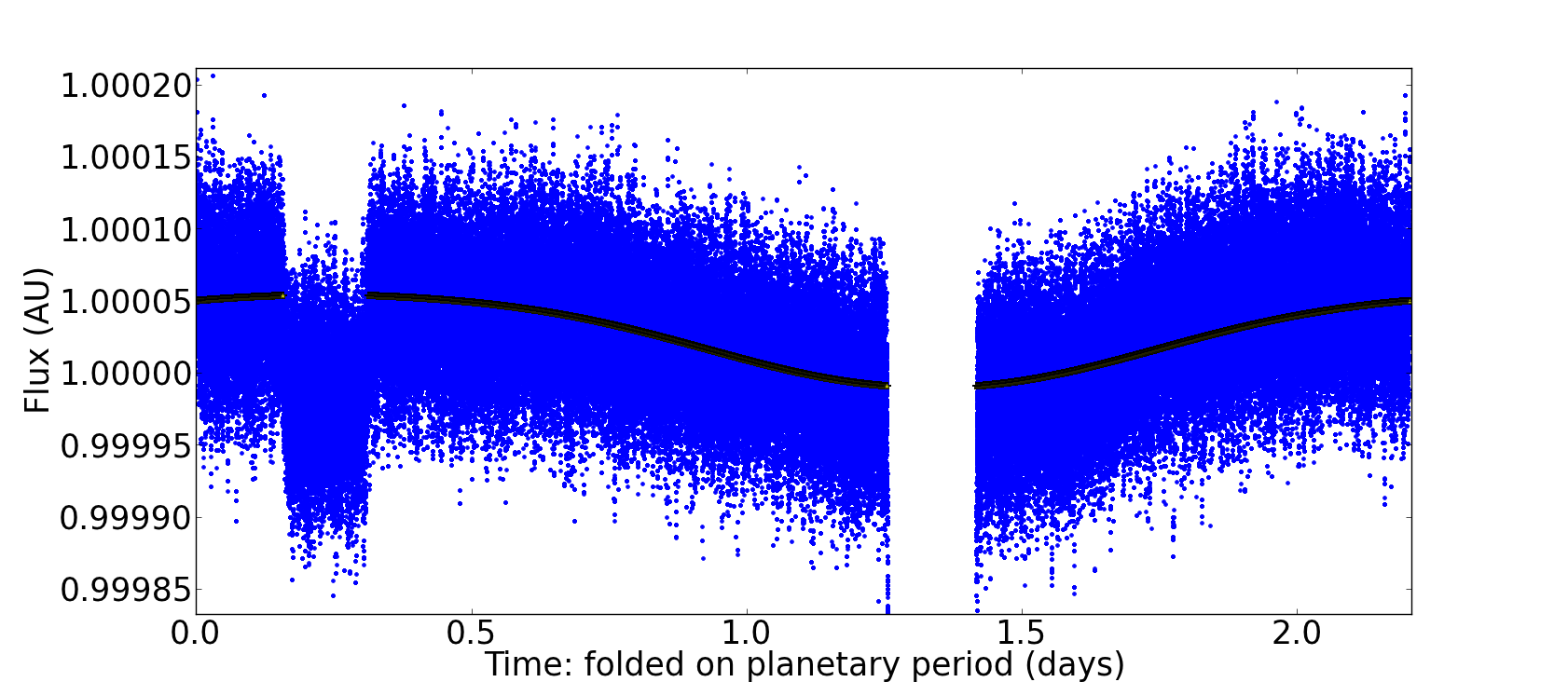}
\includegraphics[width=1\columnwidth]{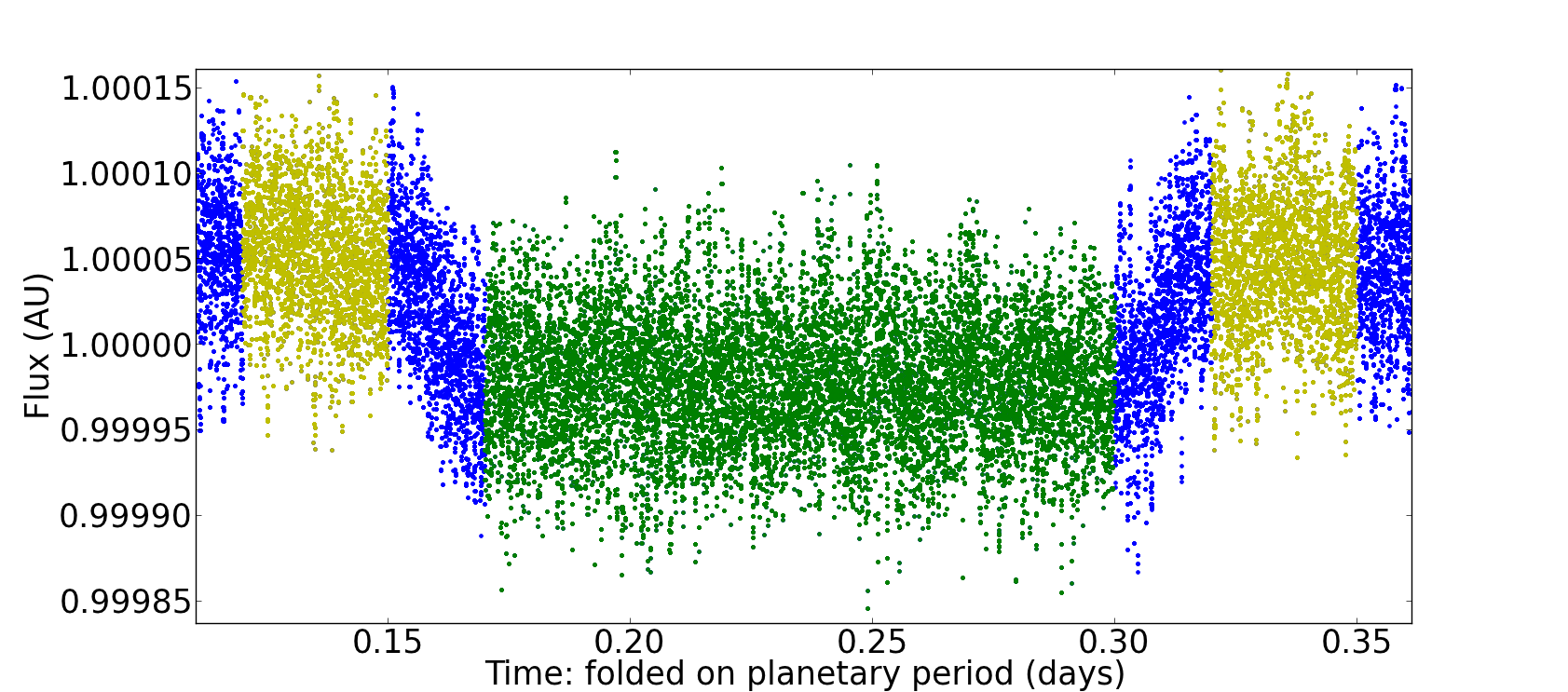}
\caption{After the planetary period is determined, the observations (blue) are folded. This allows for a detailed study of the planetary transit (top figure), phase variations (caused by planetary day-nightside and ellipsoidal variations, middle figure), and planetary occultation (bottom). The black line shows a data fit, the colored data points are used in calculating the occultation depth.
}
\label{fig:folded_flux}
\end{figure}

When the planet moves behind the star, an occultation occurs. The depth $\delta_{\rm{flux}}(\lambda)$ gives information about the planetary dayside temperature. By combining the information with the phase variations, we can also obtain the nightside temperature. We approximate both star and planet as a blackbody and use

\begin{equation}
 \delta_{\rm{flux}}(\lambda) = \left(\frac{R_{\rm p}}{R_*}\right)^2 \frac{B_\lambda (T_{\rm p})}{B_\lambda (T_\star)},
\end{equation}
where $B_\lambda (T)$ is the Planck function \citep{exoplanetsbook_chapterwinn}, which we integrate over the \textit{Kepler} bandpass. The exact occultation depth is calculated by simply comparing the averages of data points inside and just outside the occultation, as shown in Figure~\ref{fig:folded_flux}. We find an occultation depth of 71.85 $\pm$ 0.23 parts per million. We assume that all planetary flux comes from radiation (albedo zero), so the quoted temperature value is an upper limit. The indicated accuracy of the quoted temperatures should be seen as the internal error of using a black-body spectrum to describe the spectrum of the exoplanet. Since the black-body flux in the \textit{Kepler} bandpass for exoplanet temperatures will change a lot even for a small change in temperature one can estimate the black-body temperature precisely. We find the occultation to be 4.90 $\pm$ 0.25 ppm deeper than the value just outside transit.

\section{Discussion}

The asteroseismic analysis of the host star has followed largely the same methods as applied by \cite{hatp7joergen} for \textit{Kepler}'s Q0 and Q1 data. With eleven quarters now available, all parameters are determined to higher accuracy. The agreement is fairly good, but we have derived a slightly lower mass, radius and metallicity. We have not been able to model a few frequencies properly (see Figure~\ref{fig:echelle}), a problem that did not present itself in the earlier analysis by \cite{hatp7joergen} because of the fewer frequencies observed, and a lower observational accuracy. Ellipsoidal variations were found by \cite{hatp7ellipsoidalvariations}, and we confirm that the phase curve cannot be fit without including this effect. \cite{hatp7ellipsoidalvariations} find a variation of 37 ppm (no errors quoted, based on \textit{Kepler} Q0-Q1 data), which is considerably lower than our value of $59 \pm 1 {\rm \, ppm}$. All planetary values have improved accuracy. Most are in general agreement with earlier 
work. Our planetary temperatures are significantly lower than all earlier temperature determinations \citep{pal2008,hatp7science,hatp7ellipsoidalvariations,hatp7atmosphere}. They all used an occultation depth of $130 \pm 11 {\rm \, ppm}$ \citep{hatp7science}, which is almost twice as high as the value derived in this work; the analysis was done based only on \textit{Kepler} Q0 data, and the modelling did not include ellipsoidal variations as they were not yet discovered in HAT-P-7 at the time. On the other hand, the temperature in this work assumes a blackbody, which is an oversimplification and also causes the temperature error to be underestimated. \cite{hatp7ellipsoidalvariations} and \cite{hatp7atmosphere} used more realistic atmosphere models, and it should be interesting to see how they are affected by the revised occultation depth.

This work shows the excellent quality \textit{Kepler} can now deliver, both for studying stellar pulsations and planetary transits. Asteroseismology is likely to take up an increasingly important role in studying host stars of exoplanets.

\acknowledgements
The research leading to these results has received funding from the European
Research Council under the European Community's Seventh Framework Programme
(FP7/2007--2013)/ERC grant agreements n$^\circ$227224 (PROSPERITY) and n$^\circ$267864 (ASTERISK). Funding for the Stellar Astrophysics Centre is provided by The Danish National Research Foundation. We are grateful to the entire \textit{Kepler} team for the efforts leading to this highly successful mission.


\bibliography{bibtex}

\end{document}